\documentclass[prb,twocolumn,showpacs,superscriptaddress]{revtex4}

\usepackage{amsmath,amsfonts,amssymb,bm}
\usepackage{dcolumn}
\usepackage[final]{graphicx}
\usepackage{bm}

\begin{document}

\bibliographystyle{apsrev}	

\title{Gate-defined coupled quantum dots in topological insulators}

\author{Christian Ertler}
\author{Martin Raith}
\author{Jaroslav Fabian}

\affiliation{Institute for Theoretical Phyiscs,
  University of Regensburg,  93040 Regensburg, Germany}

\date{\today}

\begin{abstract}
We consider electrostatically coupled quantum dots in topological insulators, otherwise confined and gapped by a magnetic texture. By numerically solving the (2+1) Dirac equation for the wave packet dynamics, we extract 
the energy spectrum of the coupled dots as a function of bias-controlled coupling and an external 
perpendicular magnetic field.  We show that the tunneling energy can be controlled to a large extent
by the electrostatic barrier potential. Particularly interesting is the coupling via Klein tunneling through a
resonant valence state of the barrier. The effective three-level system nicely maps to a model Hamiltonian,
from which we extract the Klein coupling between the confined conduction and valence 
dots levels.  For large enough magnetic fields Klein tunneling can be completely blocked due to
the enhanced localization of the degenerate Landau levels formed in the quantum dots.

\end{abstract}

\maketitle


In topological insulators, according to the bulk-boundary correspondence principle \cite{Hasan2010:RMP, Qi2011:RMP},
topologically protected surface states are formed, which are robust against time-reversal elastic perturbations.
In the long-wavelength limit the two-dimensional (2d) electron states at the surfaces of 3D-TIs 
can be described as massless Dirac electrons with the peculiar property that the spin is locked to the momentum,
 thereby forming a helical electron gas.
Charge and spin properties become strongly intertwined, opening new 
opportunities for spintronic \cite{Zutic2004:RMP, Fabian2007:APS} applications. \cite{Raghu2010:PRL, Burkov2010:PRL, Garate2010:PRL, Yazyev2010:PRL, Yokoyama2010:PRB, Krueckl2011:PRL}

To build functional nanostructures, such as quantum dots or quantum point contacts, additional confinement
of the Dirac electrons is needed. However, conventional electrostatic confinement in a massless Dirac system is
ineffective due to Klein (interband) tunneling. In graphene this problem could be overcame by either 
mechanically cutting or etching QD-islands out of graphene flakes \cite{Ponomarenko2008:S,Guttinger2008:APS,Schnez2009:APL} or
by inducing a gap by an underlying substrate, which breaks the pseudospin symmetry. \cite{Giovannetti2007:PRB, Trauzettel2007:NatPhys}

In TIs a mass gap can be created by breaking the TR symmetry at the surface  by applying a magnetic
field. This could be achieved by proximity to a magnetic material \cite{Fu2007:PRB,Qi2008:PRB}, or by coating the surface randomly
with magnetic impurities \cite{Chen2010:S, Liu2009:PRL, Hor2010:PRB}. 
By modifying the magnetic texture of the deposited magnetic film, a spatially inhomogeneous mass
term is induced, opening the possibility to define quantum dot (QD) regions \cite{Ferreira2013:PRL},
or waveguides formed along the magnetic domain wall regions \cite{Hammer2013:APL}.
Another interesting, possibly more feasible way of defining confinement regions, 
is to induce a uniform mass gap and to define the QDs
by electrostatic gates, which are energetically shifting the band gap \cite{Recher2009:PRB, Pal2011:PRB}.
In this paper we will focus on such gate-defined topological insulator quantum dots. 

Single QDs confining Dirac electrons have been thoroughly investigated 
in the last years either by numerically solving tight-binding models or be deducing analytical solutions 
if a cylindrical symmetry and infinite-mass boundary conditions are present. \cite{Schnez2008:PRB}
However, the properties of coupled QDs are much less understood and call for a detailed, inevitably numerical
study. In this paper we investigate how an electrostatically tunable coupling strength between the dots 
and an applied external magnetic field influence the energy spectra of the double-dot system. 
The tunneling time is deduced from studying the wave packet dynamics, which needs the numerical solution of the time-dependent 
(2+1) Dirac equation. For this purpose, we use a specially developed discretization scheme, which was 
introduced and discussed in detail in Ref.~\onlinecite{Hammer2013:CPC}.
We study two different scenarios for inducing a coupling between the dots:
(i) by conventional means, i.e., by electrostatically reducing the barrier height inbetween the dots and,
(ii) by coupling the dots via Klein tunneling upon a hole state in the barrier, which can be shifted by a 
gate voltage. Especially in the second case we find a strong tunability of the coupling strength.
We also introduce toy 1d models to study tunneling of Dirac electrons analytically. Our numerical
solutions for coupled 2d quantum dots establish quantitatively strong and efficient coupling
between the dots. In the Klein tunneling regime we provide a useful three-level parametric
hopping Hamiltonian to describe the conduction and valence band couplings. Our goal is to provide
the single-electron picture of the tunneling, which could also be used as a starting point to investigate
 the Coulomb blockade physics. 

Our paper is organized as follows.
The basic idea of a gate-controlled coupling between QDs and qualitative analytical solutions 
are introduced in Sec.~\ref{sec:model}. The numerical investigation of coupled QDs is presented  
and discussed in Sec.~\ref{sec:results}.
Summary and conclusions are given in Sec.~\ref{sec:sum}.

\section{1d model of a gate confined quantum dot}\label{sec:model}

\begin{figure}[!t]
\centering
\includegraphics[width=\linewidth]{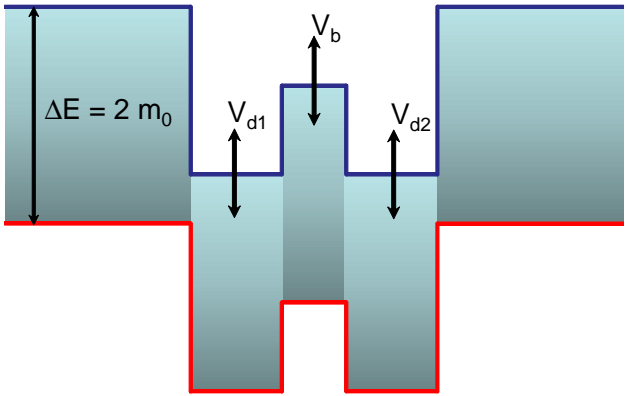}
\caption{(Color online) Schematic band profile of the conduction (blue line) and valence (red line) bands
of two coupled topological insulator quantum dots.
The uniform band gap $\Delta E = 2 m_0$ is shifted by applying gate voltages. For electrons a double dot
system is formed with the barrier height being controllable by an external bias $V_b$.
This is conventional coupling as found in semiconductor quantum dots. \cite{Stano2005:PRB}}
\label{fig:QDDscheme}
\end{figure}

\begin{figure}[!t]
\centering
\includegraphics[width=\linewidth]{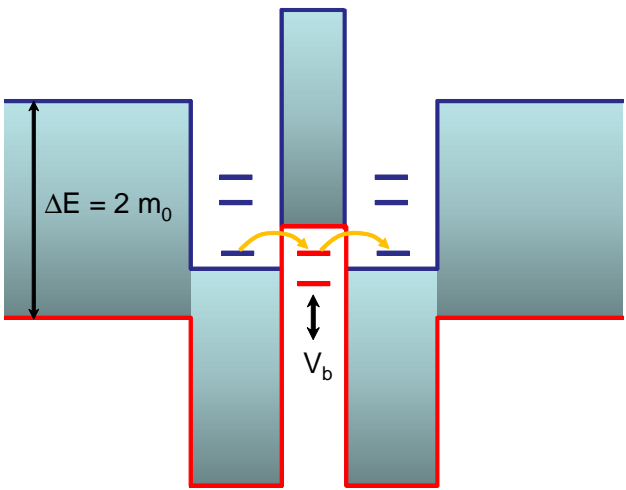}
\caption{(Color online) Schematic band profile of the conduction (blue line) and valence band (red line)
for the Klein-tunneling scenario: the coupling between the dots is realized via the Klein tunneling 
upon a hole state, which can be 
shifted by an external gate voltage $V_b$.}
\label{fig:QDDscheme2}
\end{figure}

If a uniform mass gap exists throughout the TI-surface, 
QDs can be defined by shifting the energy gap locally by applying 
gate voltages as illustrated in Fig.~\ref{fig:QDDscheme}. The mass barrier height
between the dots becomes electrostatically controllable, allowing for a direct tunability 
of the coupling strength. 
The barrier can also be shifted upwards in energy as far as
a hole state comes into resonance with the ground state of the isolated dots. This leads to 
an effective strong coupling between the dots via Klein tunneling from the electron states 
to the hole state, as illustrated in Fig.~\ref{fig:QDDscheme2}.

\begin{figure}[!t]
\centering
\includegraphics[width=\linewidth]{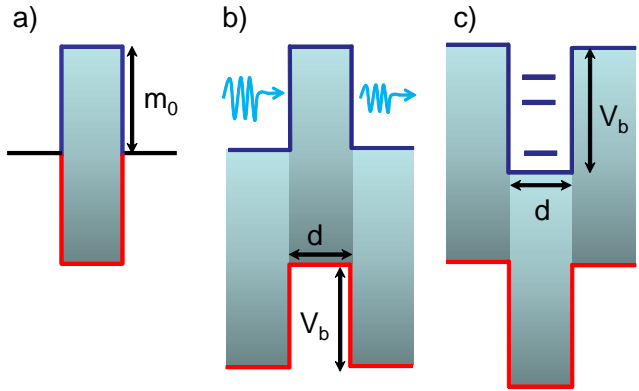}
\caption{(Color online) (a) Scheme of a single mass barrier of height $m_0$. A mass barrier (b) or a quantum well (c) of width $d$  
is formed for electrons by applying a gate voltage $V_b$  of opposite sign.}
\label{fig:QWscheme}
\end{figure}

In order to understand qualitatively how the coupling strength between QDs  
depends on the barrier height, we first investigate the transmission probability of a model
1D-mass barrier. We consider two different cases: (i) a mass barrier between leads with zero mass, as illustrated in Fig.~\ref{fig:QWscheme}~(a) 
and, (ii) a uniform mass gap in the structure with a shiftable region in the middle, 
as shown in Fig.~\ref{fig:QWscheme}~(b), which directly corresponds to the ``conventional'' coupling scenario
of Fig.~\ref{fig:QDDscheme}.

For a general 1D-structure with an inhomogeneous mass term $m(x)$ and potential $V(x)$ 
the Dirac-Hamiltonian is
\begin{equation}
H = -i\alpha\partial_x \sigma_x + m(x) \sigma_z + V(x),
\end{equation}
where $\alpha = \hbar v_f$ and $\sigma_{x,z}$ denote the Pauli matrices.
Let us assume that we can divide the region of interest into subregions in which
$m$ and $V$ can be assumed to be constant.  
For constant $m$ and $V$  the eigenfunctions  $\psi^\pm$ of left ($-$) and right ($+$) moving plane waves 
of energy $E$ are given by
\begin{equation}
\psi^{\pm} = \left(\begin{array}{c} 1\\\pm\gamma\end{array}\right) e^{\pm i q x},
\end{equation}
with 
\begin{equation}
q(V,m) = \frac{1}{\alpha}\sqrt{(E-V)^2-m^2},
\end{equation}
and 
\begin{equation}
\gamma(V,m) = \frac{\sqrt{(E-V)^2-m^2}}{(E+V)+m},
\end{equation}
yielding the general solution  $\psi = c^+\psi^+ + c^- \psi^-$.
At the boundary of neighbouring subregions $i$ and $i+1$ the
wavefunction has to be continuous, resulting in the condition
\begin{equation}
\psi_i(x_i) = \psi_{i+1}(x_i).
\end{equation}
This continuous connection of the wave functions of the subregions allows us to 
calculate the transfer matrix $M$ of the whole system, which connects the amplitudes of
the first layer $C_1 = (c^+_1,c^-_1)$ with the last, i.e., the most right one
$C_N = M C_1$.
From the elements of the transfer matrix the transmission function can then be obtained by 
\begin{equation}
T(E) = \frac{\det M}{|M_{22}|^2}.
\end{equation}

\begin{figure}[!t]
\centering
\includegraphics[width=\linewidth]{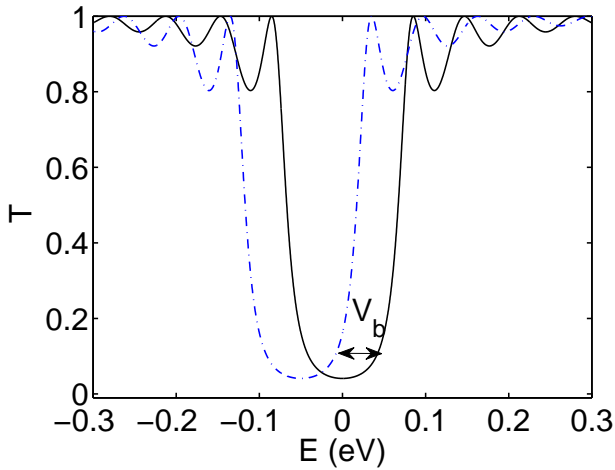}
\caption{(Color online) Calculated transmission function $T(E)$ of a single 1D-mass barrier, as shown
in Fig.~\ref{fig:QWscheme}~(a). Applying a voltage $V_b$ allows to
shift $T(E)$ along the energy axis, which strongly changes the transmission for a fixed energy.}
\label{fig:T1d}
\end{figure}

\begin{figure}[!t]
\centering
\includegraphics[width=\linewidth]{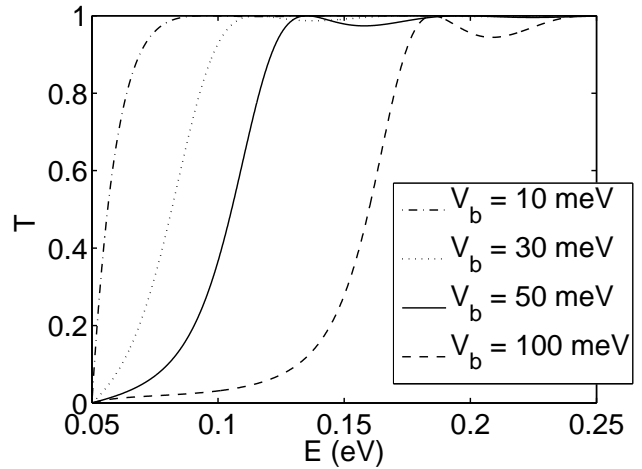}
\caption{(Color online) Calculated transmission function $T(E)$ of a gate induced 1D-barrier of width
$d = 30$~nm, as illustrated
in Fig.~\ref{fig:QWscheme}~(b), for different applied biases. The mass is set to $m_0 = 50$~meV.}
\label{fig:T1d2}
\end{figure}

\noindent {\it Mass barrier between massless dots.}
In the case (i) of a single mass barrier of height $m_0$, which is shifted by the gate voltage $V_b$
this procedure yields the
following result for the transmission function if the electron energies are below the barrier, i.e., $-m_0+V_b < E < m0+V_b$:
\begin{equation}
T(E) = 
\frac{-1+6\tilde{\gamma_b}^2-\tilde{\gamma_b}^4+
(1+\tilde{\gamma_b}^2)^2\cosh(2 d \tilde{q}_b)}
{8\tilde{\gamma_b}^2},
\end{equation}
 with $\tilde{q}_b = -i q_b$, $\tilde{\gamma}_b = -i \gamma_b$,
$\gamma_b  = \gamma(V_b,m_0)$, and
$q_b = q(V_b,m_0)$. For energies above the barrier, i.e., for $E > m_0+V_b$ and $E < -m_0+V_b$ the transmission
probability is given by
\begin{equation}
T(E) = \left[\cos^2(d q_b) + \frac{(1+\gamma_b^2)^2 \sin^2(d q_b)}{4\gamma_b^2}\right]^{-1}.
\end{equation} 
As expected one obtains an exponential and oscillatory dependence of the transmission function 
for energies smaller and greater than the
barrier height, respectively, as illustrated in Fig.~\ref{fig:T1d}. 
Applying a gate voltage allows to shift the whole transmission function along the energy axis, which
means that for a given fixed energy one obtains an exponential dependence on the applied gate voltage.
Note that in contrast to
Schr\"odinger particles the transmission remains finite even at zero energy due to the finite 
group velocity for $E=0$:
\begin{equation}
T(E=0) = \left[\cosh^2(\frac{d m_0}{\hbar v_f})\right]^{-1}\approx\mathrm{e}^
{-\frac{2 d m_0}{\hbar v_f}}.
\end{equation}

\noindent{\it Uniform mass with a gate-controlled barrier.}
In the case (ii) of a uniform mass region $m = m_0$ and a gate induced single mass barrier of height $V_b$, 
 as shown in Fig.~\ref{fig:QWscheme}~(b), the transmission function results in:
\begin{equation}
T(E) = \frac{8\gamma_l^2\gamma_b^2}{\gamma_l^4+6\gamma_l^2\gamma_b^2+\gamma_b^4-(\gamma_l^2-\gamma_b^2)^2
\cos(2d q_b)}
\end{equation}
if the electron's energy is higher than the barrier $(E > m_0+V_b)$ with $\gamma_l = \sqrt{E^2-m_0^2}/(E+m_0)$. 
For electron energies lower than the barrier $m_0 < E < m_0+V_b$ the transmission is given by
\begin{equation}
T(E) = \left\{\frac{1}{4}\left[3+\cosh(2d\tilde{q}_b)+\frac{\gamma_l^4+\tilde{\gamma}_b^4\sinh^2(d
\tilde{q}_b)}{\gamma_l^2\tilde{\gamma}_b^2}\right]\right\}^{-1}.
\end{equation}
 Again one obtains an exponential and oscillatory dependence of the transmission function 
for energies below and above the barrier height, as illustrated in Fig.~\ref{fig:T1d2} for
different barrier heights.

\noindent{\it Energy spectrum of 1d TI dots.}
Finally, we calculate the energy spectrum of a 
single dot of width $d$, as illustrated in Fig.~\ref{fig:QWscheme}~(c). By 
using the condition $\det M = 0$ the eigenenergies of the bounded states are given
by
\begin{equation}
E_n = \pm\sqrt{\left(n\frac{\pi\alpha}{d}\right)^2+m_0^2}+ V_b.
\end{equation}

\section{Coupled 2d dots}\label{sec:results}

\subsection{Numerical solution of the time-dependent 2d Dirac equation}

Here, we provide a numerical investigation of the spectrum of two coupled topological insulator two-dimensional
 quantum dots depending on
their coupling strength and external perpendicular magnetic fields $B$. 
In order to calculate the energy spectra we study the dynamics of wave packets
(see Ref.~\onlinecite{Kramer2008:JPCS} for a review of the wave packet method in general).
In comparison to a direct numerical diagonalization of the Hamiltonian, the wave packet method
allows to investigate larger systems with a higher number of grid points in a reasonable
computation time. This is needed, since the dot systems have to be large enough to 
make an effective theory actually applicable for describing the
carrier dynamics of dot systems. The energy spectrum is then obtained by a Fourier
transformation of the wave-packet autocorrelation function with the energy resolution being determined
by the total propagation time $\Delta E = 2\pi\hbar/T$. However, as a disadvantage compared to exact
diagonalization the wave packet method can miss some eigenenergy values in the case that the
corresponding amplitudes of the Fourier transformation are smaller than the numerical signal noise.

In order to obtain the energy spectrum we calculate the
local density of states 
\begin{equation}
D(E,\mathbf{r}) = -\frac{1}{\pi}\mathrm{Im}\left [G(\mathbf{r},\mathbf{r};E)\right ],
\end{equation}
which is defined via the diagonal elements of the retarded Green's function $G(E)$.
Based on the dynamics of a single wave packet initially centered at $\mathbf{r}$ 
the retarded Green' function can be constructed from 
its autocorrelation function $C(t)$
\begin{equation}\label{eq:Gret}
G(\mathbf{r},\mathbf{r};E) \approx \frac{1}{i\hbar}\int_0^\infty\mathrm{d} t 
\mathrm{e}^{i E t/\hbar} C(t),
\end{equation}
with
\begin{equation}
C(t) = \int\mathrm{d}\mathbf{r} \psi(\mathbf{r},0)^*\psi(\mathbf{r},t).
\end{equation}
Eq.~(\ref{eq:Gret}) becomes exact for a $\delta$-distributed initial state, whereas in
the numerical simulations a Gaussian shaped initial state is used \cite{Kramer2008:JPCS}.
To obtain the correlation function one has to keep track of the wave packet transient,
which requires the solution of the time dependent 2D-Dirac equation
\begin{equation}
\frac{\partial\psi(\mathbf{r},t)}{\partial t} = H_D \psi(\mathbf{r},t).
\end{equation}
The single-particle Hamiltonian at the surface of TIs can be derived within an effective field theory 
approach
\cite{Hasan2010:RMP,Qi2011:RMP}, yielding 
\begin{equation}
H_D = v_f\left\{ \left [ -i\hbar\nabla-e\mathbf{A}(\mathbf{r},t) \right ] \times\hat{z}\right\}\cdot\bm{\sigma}
+m_z(\mathbf{r})\sigma_z-e\phi(\mathbf{r},t),
\end{equation}
where $\mathbf{A}(\mathbf{r},t)$ and $\phi(\mathbf{r},t)$
denotes the space- and time-dependent vector and electrostatic
potential, respectively. 
The inhomogeneous mass term $m_z(\mathbf{r})\sigma_z$ is induced by 
breaking the TR symmetry at the TI-surface, e.g., by 
proximity of a magnetic layer \cite{Fu2007:PRB,Qi2008:PRB} or by magnetic doping \cite{Chen2010:S, Liu2009:PRL,
Hor2010:PRB}.    
The order of magnitude for the mass-gap $\Delta_\mathrm{gap} = 2 |m_z|$ can be expected to be tens of meV.

In order to solve numerically the (2+1) Dirac equation we use a specially developed staggered-grid leap 
frog scheme, which we introduced and discussed in detail in Ref.~\onlinecite{Hammer2013:CPC}.
The numerical solution of the Dirac
equation on a finite grid is a more subtle issue than for the
non-relativistic Schr{\"o}dinger equation. As well known from
lattice field-theory, discretization of the Dirac equation leads to
the so-called fermion-doubling problem, i.e., for large wavevectors a
wrong energy dispersion is revealed. This leads to the doubling of the
eigenstates at a fixed energy value. For a longtime
propagation it is of great importance to use an almost
dispersion-preserving finite-difference scheme, since scattering at
spatiotemporal potentials and at the boundary can introduce higher
wavevector components even when one starts with a wavepacket with
its wavevector-components closely centered at $k = 0$. Moreover,
the formulation of proper boundary conditions is crucial to avoid
spurious reflections and eventually instability \cite{Hammer2013:CPC}.

\subsection{Energy spectra of single and coupled QDs}

\subsubsection{Energy spectra of single QDs}

\begin{figure}[!t]
\centering
\includegraphics[width=\linewidth]{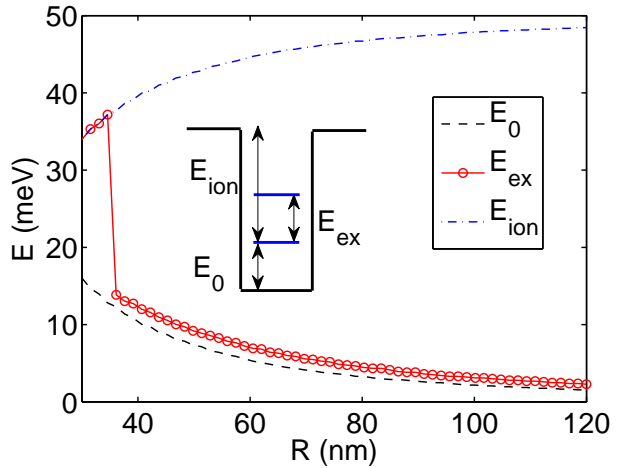}
\caption{(Color online) The ground state energy $E_0$ (solid line), 
the excitation energy $E_\mathrm{ex}$ (dashed line), and the ionization
energy $E_\mathrm{ion}$ (dash-dotted line) of single dot for different radii $R$ at zero
magnetic field.}
\label{fig:QDRepsis}
\end{figure}

\begin{figure}[!t]
\centering
\includegraphics[width=\linewidth]{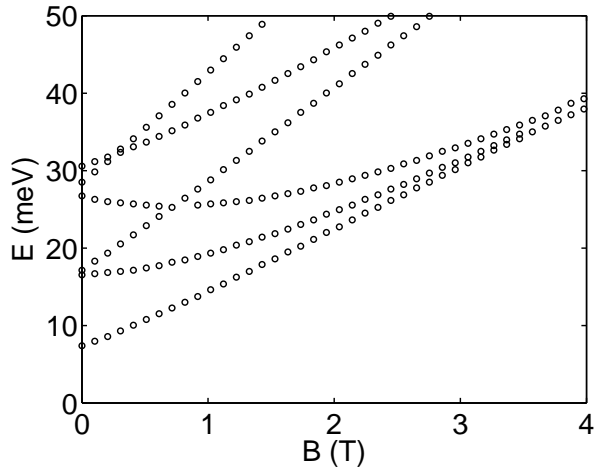}
\caption{(Color online) Magnetic field ($B$) dependent energy spectrum of a single QD of radius $R = 50$~nm.
The energy levels converge to Landau levels for higher $B$-fields.}
\label{fig:QDB}
\end{figure}

\begin{figure}[!t]
\centering
\includegraphics[width=\linewidth]{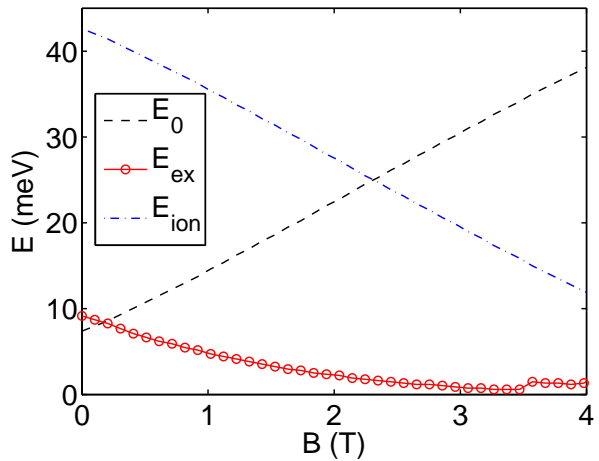}
\caption{(Color online) Magnetic field ($B$) dependence of the ground state energy $E_0$ (solid line), 
the excitation energy $E_{ex}$ (dashed line), and the ionization
energy $E_{ion}$ (dash-dotted line) of single dot of radius $R = 50$~nm.}
\label{fig:QDBepsis}
\end{figure}

First, we investigate the confinement energy of a single isolated QD as a function of 
radius $R$ and the magnetic field $B$. The circular dot potential $\phi_d(r_0)$ centered at $r_0$   
is assumed to be described by Fermi-Dirac function 
\begin{equation}
 \phi_d(r_0) = V_d\tilde{\phi}_d(r_0)= V_d F_D(r_0-r),
\end{equation}
with  $F_D(x) = [1+\exp(x/\beta_r)]^{-1}$; $V_d$  denotes the potential height. The potential step 
is smeared on the range of $\beta_r = 0.01 R$. In the following we set the Fermi velocity to $v_f = 10^5$~m/s 
and the dot potential height is chosen as $|V_d| = 50$~meV.
Figure~\ref{fig:QDRepsis} shows the ground state or confinement energy $E_0$ of the dot,
the excitation energy defined as difference of the ground and first excited dot state
$E_{\mathrm{ex}}= E_1 - E_0$, and the ionization energy $E_{\mathrm{ion}} = |V_d| - E_1 $, as given by
the energy difference of the ground state energy to the continuum of the delocalized states, as a function
of different dot radii at zero magnetic field. 
For radii smaller than about $R < 38$~nm only a single bound state exists in the QD and, hence, 
$E_{\mathrm{ex}} = E_{\mathrm{ion}}$. As expected, the weaker confinement of the Dirac electrons in larger
dot systems leads to a decreasing of both the ground state energy $E_0$ and the excitation energy 
$E_{\mathrm{ex}}$, as shown in Fig.~\ref{fig:QDRepsis}. Note, that the typical confinement energy of
Dirac electrons is of the order of 10~meV, which is an order of magnitude higher than in conventional
semiconductor dots of comparable size.
The energy spectrum for the lowest QD-levels versus magnetic field, which is applied perpendicular to the
TI-surface, is plotted in Fig.~\ref{fig:QDB} assuming a fixed radius of $R = 50$~nm. 
A detailed analytical study of the energy spectrum of a single graphene
quantum dot in a perpendicular magnetic field is given in Ref.~\onlinecite{Schnez2008:PRB}. The eigenspectrum of 
the dot for $B = 0$ is obtained by solving 
the implicit equation 
\begin{equation}
J_m(k R) = J_{m+1}( k R)
\end{equation}
with $J_m$ denoting the Bessel functions of first kind of order $m$ and $ E = \hbar v_f k$. Since the 
total angular momentum $J_z = l_z+(\hbar/2)\sigma_z$, with
$l_z$ denoting the orbital momentum operator, commutes with the Hamiltonian $([H,J_z] = 0)$, $m$ is a good
quantum number. Hence, the ground state with $l = 0$ is doubly degenerate according to its spin. 
For higher magnetic fields
the levels start to converge to degenerate Landau levels, which are determined by the expression \cite{Schnez2008:PRB}
\begin{equation}
E_m = v_f\sqrt{2 e \hbar B (m+1)}.
\end{equation}
As can be seen in Fig.~\ref{fig:QDB} at about $B \approx 3.6$~T the first two levels converge numerically, 
which leads to a sudden kink in the magnetic field
dependence of the excitation energy $E_\mathrm{ex}$, as indicated by the solid red line 
in Fig.~\ref{fig:QDBepsis}. The magnetic field effectively 
acts as an additional confinement causing an almost linear enhancement of the confinement energy $E_0$. 

\subsubsection{Energy spectra of coupled QDs}

\begin{figure}[!t]
\centering
\includegraphics[width=\linewidth]{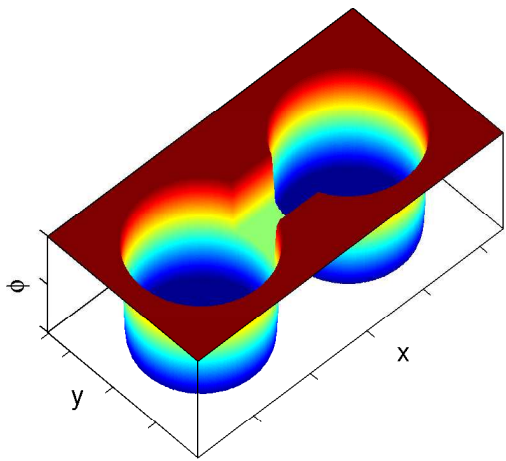}
\caption{(Color online) The double dot potential in the TI-surface. The coupling of the two dots is controlled
by shifting the barrier potential.}
\label{fig:QDD3d}
\end{figure}

Our double dot system comprises two circular disks, which are connected by a potential bridge, as 
shown in Fig.~\ref{fig:QDD3d}.
The barrier or bridge potential $\phi_b = V_b\tilde{\phi}_b$ is described by a rectangular step-function of width $w$ and
length $d$, which
is smeared at the boundaries by Fermi-Dirac function 
\begin{equation}
 \tilde{\phi}_b = F_D(-w/2-y)F_D(y-w/2)F_D(-d-x)F_D(x-d)
\end{equation}
The total potential of the coupled QD-system can then be defined by
\begin{equation}
 \phi(\mathbf{r}) = V_d \tilde{\phi}_{dd}+
V_b \max(|\tilde{\phi}_b|- |\tilde{\phi}_{dd}|,0)
\end{equation}
with 
$\tilde{\phi}_{dd}= \max[\tilde{\phi}_d(r_1),\tilde{\phi}_d(r_2)$] describing
the potential of the decoupled double dot system.

For the quantitative simulations we choose the following structure parameters:
dot radius $R = 50$~nm, bridge length $d = 30$~nm, bridge width $w = 40$~nm, grid resolution
$\Delta_x =\Delta_y=1$~nm, and
an uniform mass term of $m_0 = 50$~meV. To ensure the stability of the discretization scheme \cite{Hammer2013:CPC} the
Courant-Friedrichs-Lewy (CFL) condition has to be fulfilled $\Delta t < \min(\Delta_x,\Delta_y)/v_f$. We use typically $N_t = 1.6\times 10^5$ time steps for each
wave packet propagation, which leads to an energy resolution of about $\Delta_E\approx 0.1$~meV in the 
discrete Fast Fourier transformation (FFT), where $\Delta_E = 2\pi\hbar/(N_t\Delta_t)$. The finite energy 
resolution of the fixed energy grid
of the discrete FFT becomes noticeable in the following plots of the energy spectrum as small discontinuous 
jumps when external parameters, such as the barrier voltage, are changed.

\begin{figure}[!t]
\centering
\includegraphics[width=\linewidth]{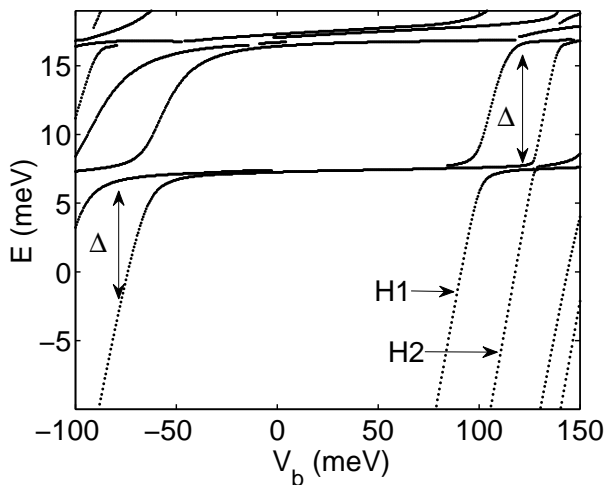}
\caption{(Color online) Calculated energy spectrum of the double TI QD system versus applied barrier gate voltage for
$B = 0$.}
\label{fig:QDD1}
\end{figure}

\begin{figure}[!t]
\centering
\includegraphics[width=\linewidth]{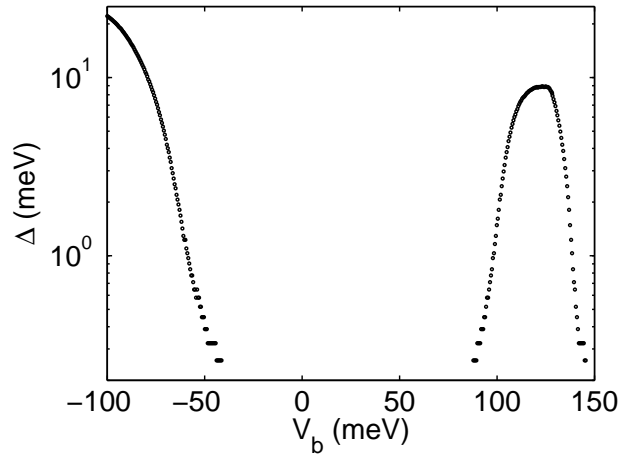}
\caption{(Color online) Energy splitting $\Delta$ of the first bonding and antibonding state (as extracted from Fig.~\ref{fig:QDD1}) versus the applied bias $V_b$.}
\label{fig:Deps}
\end{figure}

\begin{figure}[!t]
\centering
\includegraphics[width=\linewidth]{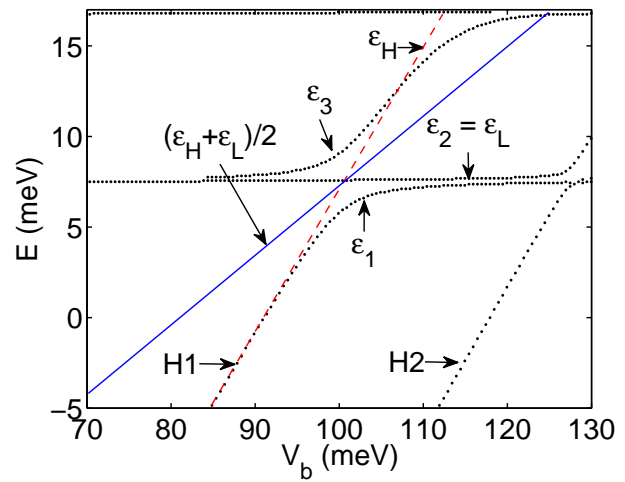}
\caption{(Color online) Zoom-in of Fig.~\ref{fig:QDD1} in the region, where the two dots are coupled
via Klein tunneling upon approaching hole state $H_1$. An effective three-level model reveals that the
anticrossing should be symmetric around $(\varepsilon_L+\varepsilon_H)/2$.}
\label{fig:QDD1a}
\end{figure}

\begin{figure}[!t]
\centering
\includegraphics[width=\linewidth]{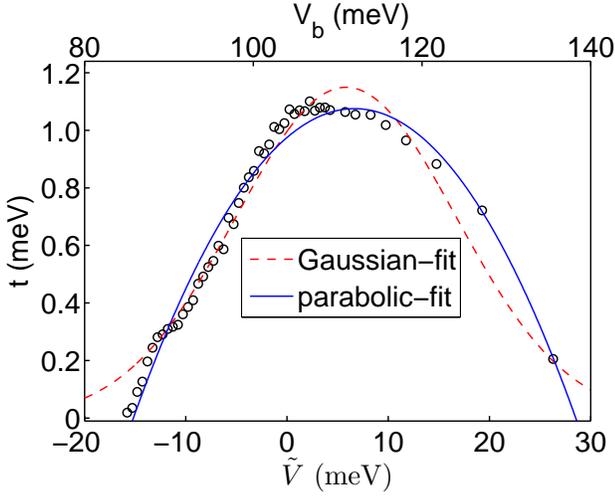}
\caption{(Color online) Voltage-dependence of the hopping parameter $t$ of the effective three-level model as extracted from
the numerical data of the first eigenvalue $\varepsilon_1$.}
\label{fig:tfit}
\end{figure}

The dependence of the energy spectrum on the gate-controlled barrier height $V_b$ for $B = 0$ is shown
in Fig.~\ref{fig:QDD1}. For $V_b$ = 0 the two dots are almost isolated. To realize the
``conventional coupling'' as illustrated in Fig.~\ref{fig:QDDscheme} a negative
bias has to be applied, which reduces the barrier potential. At around $V_b = -40$~meV the bonding and antibonding states 
start to be split in energy by $\Delta$ due to the increasing coupling between the dots.
From the energy splitting the tunneling time follows as $\tau =2\pi\hbar/\Delta$.

If a positive bias is applied, the hole states in the barrier region are shifted upwards in energy enabling 
at some point the electrons to hop by Klein tunneling from one QD to the other via the hole state, as illustrated
in Fig.~\ref{fig:QDDscheme2}. The hybridization of the two electron levels and the hole state induces an 
anticrossing of the first excited electron state and the hole state, leading to a strongly tunable
excitation energy  of maximally $\Delta \approx 8$~meV, giving a typical tunneling time of $\tau \approx 0.5$~ps.    
Figure~\ref{fig:QDD1a} shows this anticrossing as a zoom-in of Fig.~\ref{fig:QDD1}.

The main features of Fig.~\ref{fig:QDD1a} can be understand by an effective three-level model.
As a starting point we assume that the direct hopping between the
 left and right single dot ground states $|L\rangle$ and $|R\rangle$, respectively, is inhibited
and only hopping via the hole state $|H\rangle$ is possible. Then the effective Hamiltonian in the
basis $\{|L\rangle, |H\rangle, |R\rangle\}$ reads 
\begin{eqnarray}
H_0 = \begin{pmatrix}
\varepsilon_L & it & 0\\
-it^* & \varepsilon_H & it\\
0 & -it^* & \varepsilon_L \end{pmatrix}
\end{eqnarray}
with $\varepsilon_R = \varepsilon_L$, and $t$ denotes the hopping amplitude.
The eigenenergies are given by
\begin{equation}\label{eigenenergies}
\begin{aligned}
\varepsilon_2 &= \varepsilon_L, ~ \text{and} \\
\varepsilon_{1/3} &= \frac{\varepsilon_L+\varepsilon_H}{2} \mp\frac{1}{2}\sqrt{(
\varepsilon_L-\varepsilon_H)^2+8t^2},
\end{aligned}
\end{equation}
and the (unnormalized) eigenstates result in 
\begin{equation}\label{eigenstates}
\begin{aligned}
\varphi_2^{(0)} &= (1,0,1), ~ \text{and} \\
\varphi_{1/3}^{(0)} &=(-1,\xi\left(-i\mp\sqrt{1+2/\xi^2}\right),1) \\
\end{aligned}
\end{equation}
with $\xi = (\varepsilon_L-\varepsilon_H)/2t$. This suggests that one eigenenergy value should
remain almost unaffected and that 
the anticrossing should be symmetric around $(\varepsilon_L+\varepsilon_H)/2$.
As illustrated in Fig.~\ref{fig:QDD1a} this behaviour is approximately fulfilled by our numerical
simulation results.

If one introduces  an additional weak direct coupling between the dots as described by the Hamiltonian  
\begin{eqnarray}
H_1 = \begin{pmatrix}
0 & 0 & it_1\\
0 & 0 & 0\\
-it_1 & 0 & 0 \end{pmatrix},
\end{eqnarray}
perturbation theory yields that the eigenergies of $H_0$ are not changed in first order.
However, now a small component of the hole state of the order of $t_1/(\varepsilon_2-\varepsilon_1)$
mixes to the eigenstate of 
$\varepsilon_2 = \varepsilon_L$:
\begin{equation}
|\varphi_2\rangle = |\varphi_2\rangle^{(0)}-\frac{2it_1}{\varepsilon_2-\varepsilon_1}
 |\varphi_1\rangle^{(0)}-\frac{2it_1}{\varepsilon_2-\varepsilon_3}|\varphi_3\rangle^{(0)}.
 \end{equation}

\begin{figure}[!t]
\centering
\includegraphics[width=\linewidth]{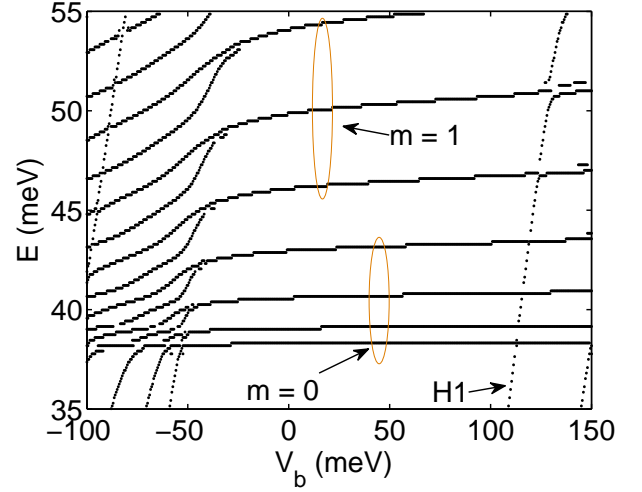}
\caption{(Color online) Energy spectrum of the double dot system versus applied barrier gate voltage for
$B = 4$~T.}
\label{fig:QDD4}
\end{figure}

From our numerical data we can extract the voltage-dependence of the hopping parameter $t$.
Therefore, we redefine the origin of the coordinate system as the crossing point in which $\varepsilon_L = 
\varepsilon_H$, i.e., at the point $\varepsilon_L = 7.5$~meV and $V_0 = 100.82$~meV.
Then $\varepsilon_L = 0$ by definition and the bias-dependent hole state is described by the
asymptotic linear function $\varepsilon_H(V_b) = k_H (V_b-V_0) = k_H \tilde{V}$, shown as
dashed (red) line in Fig.~\ref{fig:QDD1a}, with $k_H = 0.77$ being obtained by linear regression. 
From the numerical results for the lowest eigenvalue $\varepsilon_1(\tilde{V})$ we calculate the hopping parameter, 
which is given by
\begin{equation}
t(\tilde{V}) = \sqrt{\frac{\varepsilon_1^2-\varepsilon_1\varepsilon_H}{2}}.
\end{equation}
Figure~\ref{fig:tfit} shows the obtained voltage-dependence of the hopping parameter. Since the coupling between
the electron and hole state vanishes for large voltages, i.e.,
$\lim_{\tilde{V}\rightarrow\pm \infty} t(\tilde{V}) = 0$, we fit our numerical results to an Gaussian, as shown in
Fig.~\ref{fig:tfit}. For comparison also a parabolic fit is provided. The effective model is most suitable in the region
for $\tilde{V}$ between 0 and 10 meV, where $t \approx 1$~meV is roughly constant. Otherwise the model
is to be treated as a convenient parametric fit.


A qualitatively different dependence of the energy spectrum on the applied barrier bias $V_b$ is found
 if a strong enough magnetic field is applied, which induces the formation of Landau levels of 
 magnetic quantum numbers $m$ corresponding to the total angular momentum $J_z = l_z+\hbar/2\sigma_z$ with
$l_z$ denoting the orbital momentum. 
As shown in Fig.~\ref{fig:QDD4} (for $B = 4$~T) the first hole level almost does not couple to the energy levels of
the first Landau niveaus with $m = 0$. This follows from the fact that due to the $B$-field 
the states are more localized, so that their overlap, on which the hopping parameter essentially depends, is exponentially smaller. This is most pronounced for the lowest state for $m = 0$ but can be also seen for the first excited states ($m = 1$), where
 the effect of anticrossing is smaller than for the case of a vanishing magnetic field.

\section{Conclusions}\label{sec:sum}

We have shown that the coupling between two quantum dots, which are geometrically defined by gate electrodes, 
can be strongly modulated and controlled by both a gate bias, which brings a hole level into resonance with the
electron states, and a perpendicular magnetic field, which changes the symmetry properties of the 
confined states. The anticrossing of the hole state and the dot ground states can be qualitatively understood within
a three-level model, in which only hopping to the hole state is assumed.
The Klein-tunneling assisted coupling leads to energy splittings of the order of
10~meV, which corresponds to typical tunneling times of several hundreds of femtoseconds.

\section{Acknowledgment}

This work has been supported by the German Research Foundation under the grants SFB 689 and GRK 1570.
We thank Peter Stano for very valuable and fruitful discussions.

\bibliography{ti}

\begin{thebibliography}{28}
\expandafter\ifx\csname natexlab\endcsname\relax\def\natexlab#1{#1}\fi
\expandafter\ifx\csname bibnamefont\endcsname\relax
  \def\bibnamefont#1{#1}\fi
\expandafter\ifx\csname bibfnamefont\endcsname\relax
  \def\bibfnamefont#1{#1}\fi
\expandafter\ifx\csname citenamefont\endcsname\relax
  \def\citenamefont#1{#1}\fi
\expandafter\ifx\csname url\endcsname\relax
  \def\url#1{\texttt{#1}}\fi
\expandafter\ifx\csname urlprefix\endcsname\relax\def\urlprefix{URL }\fi
\providecommand{\bibinfo}[2]{#2}
\providecommand{\eprint}[2][]{\url{#2}}

\bibitem[{\citenamefont{Hasan and Kane}(2010)}]{Hasan2010:RMP}
\bibinfo{author}{\bibfnamefont{M.~Z.} \bibnamefont{Hasan}} \bibnamefont{and}
  \bibinfo{author}{\bibfnamefont{C.~L.} \bibnamefont{Kane}}
  (\bibinfo{year}{2010}).

\bibitem[{\citenamefont{Qi and Zhang}(2011)}]{Qi2011:RMP}
\bibinfo{author}{\bibfnamefont{X.-L.} \bibnamefont{Qi}} \bibnamefont{and}
  \bibinfo{author}{\bibfnamefont{S.-C.} \bibnamefont{Zhang}},
  \bibinfo{journal}{Rev. Mod. Phys.} \textbf{\bibinfo{volume}{83}},
  \bibinfo{pages}{1057} (\bibinfo{year}{2011}).

\bibitem[{\citenamefont{{\v{Z}uti\'c} et~al.}(2004)\citenamefont{{\v{Z}uti\'c},
  Fabian, and {Das Sarma}}}]{Zutic2004:RMP}
\bibinfo{author}{\bibfnamefont{I.}~\bibnamefont{{\v{Z}uti\'c}}},
  \bibinfo{author}{\bibfnamefont{J.}~\bibnamefont{Fabian}}, \bibnamefont{and}
  \bibinfo{author}{\bibfnamefont{S.}~\bibnamefont{{Das Sarma}}},
  \bibinfo{journal}{Rev. Mod. Phys.} \textbf{\bibinfo{volume}{76}},
  \bibinfo{pages}{323} (\bibinfo{year}{2004}).

\bibitem[{\citenamefont{Fabian et~al.}(2007)\citenamefont{Fabian,
  Matos-Abiague, Ertler, Stano, and \v{Z}uti\'c}}]{Fabian2007:APS}
\bibinfo{author}{\bibfnamefont{J.}~\bibnamefont{Fabian}},
  \bibinfo{author}{\bibfnamefont{A.}~\bibnamefont{Matos-Abiague}},
  \bibinfo{author}{\bibfnamefont{C.}~\bibnamefont{Ertler}},
  \bibinfo{author}{\bibfnamefont{P.}~\bibnamefont{Stano}}, \bibnamefont{and}
  \bibinfo{author}{\bibfnamefont{I.}~\bibnamefont{\v{Z}uti\'c}},
  \bibinfo{journal}{Acta Phys. Slovaca} \textbf{\bibinfo{volume}{57}},
  \bibinfo{pages}{565} (\bibinfo{year}{2007}).

\bibitem[{\citenamefont{Raghu et~al.}(2010)\citenamefont{Raghu, Chung, Qi, and
  Zhang}}]{Raghu2010:PRL}
\bibinfo{author}{\bibfnamefont{S.}~\bibnamefont{Raghu}},
  \bibinfo{author}{\bibfnamefont{S.~B.} \bibnamefont{Chung}},
  \bibinfo{author}{\bibfnamefont{X.-L.} \bibnamefont{Qi}}, \bibnamefont{and}
  \bibinfo{author}{\bibfnamefont{S.-C.} \bibnamefont{Zhang}},
  \bibinfo{journal}{Phys. Rev. Lett.} \textbf{\bibinfo{volume}{104}}
  (\bibinfo{year}{2010}).

\bibitem[{\citenamefont{Burkov and Hawthorn}(2010)}]{Burkov2010:PRL}
\bibinfo{author}{\bibfnamefont{A.~A.} \bibnamefont{Burkov}} \bibnamefont{and}
  \bibinfo{author}{\bibfnamefont{D.~G.} \bibnamefont{Hawthorn}},
  \bibinfo{journal}{Phys. Rev. Lett.} \textbf{\bibinfo{volume}{105}}
  (\bibinfo{year}{2010}).

\bibitem[{\citenamefont{Garate and Franz}(2010)}]{Garate2010:PRL}
\bibinfo{author}{\bibfnamefont{I.}~\bibnamefont{Garate}} \bibnamefont{and}
  \bibinfo{author}{\bibfnamefont{M.}~\bibnamefont{Franz}}
  (\bibinfo{year}{2010}).

\bibitem[{\citenamefont{Yazyev et~al.}(2010)\citenamefont{Yazyev, Moore, and
  Louie}}]{Yazyev2010:PRL}
\bibinfo{author}{\bibfnamefont{O.~V.} \bibnamefont{Yazyev}},
  \bibinfo{author}{\bibfnamefont{J.~E.} \bibnamefont{Moore}}, \bibnamefont{and}
  \bibinfo{author}{\bibfnamefont{S.~G.} \bibnamefont{Louie}}
  (\bibinfo{year}{2010}).

\bibitem[{\citenamefont{Yokoyama et~al.}(2010)\citenamefont{Yokoyama, Tanaka,
  and Nagaosa}}]{Yokoyama2010:PRB}
\bibinfo{author}{\bibfnamefont{T.}~\bibnamefont{Yokoyama}},
  \bibinfo{author}{\bibfnamefont{Y.}~\bibnamefont{Tanaka}}, \bibnamefont{and}
  \bibinfo{author}{\bibfnamefont{N.}~\bibnamefont{Nagaosa}}
  (\bibinfo{year}{2010}).

\bibitem[{\citenamefont{Krueckl and Richter}(2011)}]{Krueckl2011:PRL}
\bibinfo{author}{\bibfnamefont{V.}~\bibnamefont{Krueckl}} \bibnamefont{and}
  \bibinfo{author}{\bibfnamefont{K.}~\bibnamefont{Richter}},
  \bibinfo{journal}{Phys. Rev. Lett.} \textbf{\bibinfo{volume}{107}}
  (\bibinfo{year}{2011}).

\bibitem[{\citenamefont{Ponomarenko et~al.}(2008)\citenamefont{Ponomarenko,
  Schedin, Katsnelson, Yang, Hill, Novoselov, and Geim}}]{Ponomarenko2008:S}
\bibinfo{author}{\bibfnamefont{L.~A.} \bibnamefont{Ponomarenko}},
  \bibinfo{author}{\bibfnamefont{F.}~\bibnamefont{Schedin}},
  \bibinfo{author}{\bibfnamefont{M.~I.} \bibnamefont{Katsnelson}},
  \bibinfo{author}{\bibfnamefont{R.}~\bibnamefont{Yang}},
  \bibinfo{author}{\bibfnamefont{E.~W.} \bibnamefont{Hill}},
  \bibinfo{author}{\bibfnamefont{K.~S.} \bibnamefont{Novoselov}},
  \bibnamefont{and} \bibinfo{author}{\bibfnamefont{A.~K.} \bibnamefont{Geim}},
  \textbf{\bibinfo{volume}{320}}, \bibinfo{pages}{356} (\bibinfo{year}{2008}).

\bibitem[{\citenamefont{Guttinger et~al.}(2008)\citenamefont{Guttinger,
  Stampfer, Hellmuller, Molitor, Ihn, and Ensslin}}]{Guttinger2008:APS}
\bibinfo{author}{\bibfnamefont{J.}~\bibnamefont{Guttinger}},
  \bibinfo{author}{\bibfnamefont{C.}~\bibnamefont{Stampfer}},
  \bibinfo{author}{\bibfnamefont{S.}~\bibnamefont{Hellmuller}},
  \bibinfo{author}{\bibfnamefont{F.}~\bibnamefont{Molitor}},
  \bibinfo{author}{\bibfnamefont{T.}~\bibnamefont{Ihn}}, \bibnamefont{and}
  \bibinfo{author}{\bibfnamefont{K.}~\bibnamefont{Ensslin}},
  \bibinfo{journal}{Applied Physics Letters} \textbf{\bibinfo{volume}{93}},
  \bibinfo{eid}{212102} (pages~\bibinfo{numpages}{3}) (\bibinfo{year}{2008}).

\bibitem[{\citenamefont{Schnez et~al.}(2009)\citenamefont{Schnez, Molitor,
  Stampfer, Guttinger, Shorubalko, Ihn, and Ensslin}}]{Schnez2009:APL}
\bibinfo{author}{\bibfnamefont{S.}~\bibnamefont{Schnez}},
  \bibinfo{author}{\bibfnamefont{F.}~\bibnamefont{Molitor}},
  \bibinfo{author}{\bibfnamefont{C.}~\bibnamefont{Stampfer}},
  \bibinfo{author}{\bibfnamefont{J.}~\bibnamefont{Guttinger}},
  \bibinfo{author}{\bibfnamefont{I.}~\bibnamefont{Shorubalko}},
  \bibinfo{author}{\bibfnamefont{T.}~\bibnamefont{Ihn}}, \bibnamefont{and}
  \bibinfo{author}{\bibfnamefont{K.}~\bibnamefont{Ensslin}},
  \bibinfo{journal}{Applied Physics Letters} \textbf{\bibinfo{volume}{94}},
  \bibinfo{eid}{012107} (pages~\bibinfo{numpages}{3}) (\bibinfo{year}{2009}).

\bibitem[{\citenamefont{Giovannetti et~al.}(2007)\citenamefont{Giovannetti,
  Khomyakov, Brocks, Kelly, and van~den Brink}}]{Giovannetti2007:PRB}
\bibinfo{author}{\bibfnamefont{G.}~\bibnamefont{Giovannetti}},
  \bibinfo{author}{\bibfnamefont{P.~A.} \bibnamefont{Khomyakov}},
  \bibinfo{author}{\bibfnamefont{G.}~\bibnamefont{Brocks}},
  \bibinfo{author}{\bibfnamefont{P.~J.} \bibnamefont{Kelly}}, \bibnamefont{and}
  \bibinfo{author}{\bibfnamefont{J.}~\bibnamefont{van~den Brink}},
  \bibinfo{journal}{Phys. Rev. B} \textbf{\bibinfo{volume}{76}},
  \bibinfo{pages}{073103} (\bibinfo{year}{2007}).

\bibitem[{\citenamefont{Trauzettel et~al.}(2007)\citenamefont{Trauzettel,
  Bulaev, Loss, and Burkard}}]{Trauzettel2007:NatPhys}
\bibinfo{author}{\bibfnamefont{B.}~\bibnamefont{Trauzettel}},
  \bibinfo{author}{\bibfnamefont{D.~V.} \bibnamefont{Bulaev}},
  \bibinfo{author}{\bibfnamefont{D.}~\bibnamefont{Loss}}, \bibnamefont{and}
  \bibinfo{author}{\bibfnamefont{G.}~\bibnamefont{Burkard}},
  \bibinfo{journal}{Nat. Phys.} \textbf{\bibinfo{volume}{3}},
  \bibinfo{pages}{1745} (\bibinfo{year}{2007}).

\bibitem[{\citenamefont{Fu and Kane}(2007)}]{Fu2007:PRB}
\bibinfo{author}{\bibfnamefont{L.}~\bibnamefont{Fu}} \bibnamefont{and}
  \bibinfo{author}{\bibfnamefont{C.~L.} \bibnamefont{Kane}}
  (\bibinfo{year}{2007}).

\bibitem[{\citenamefont{Qi et~al.}(2008)\citenamefont{Qi, Hughes, and
  Zhang}}]{Qi2008:PRB}
\bibinfo{author}{\bibfnamefont{X.-L.} \bibnamefont{Qi}},
  \bibinfo{author}{\bibfnamefont{T.~L.} \bibnamefont{Hughes}},
  \bibnamefont{and} \bibinfo{author}{\bibfnamefont{S.-C.} \bibnamefont{Zhang}}
  (\bibinfo{year}{2008}).

\bibitem[{\citenamefont{Chen et~al.}(2010)\citenamefont{Chen, Chu, Analytis,
  Liu, Igarashi, Kuo, Qi, Mo, Moore, Lu et~al.}}]{Chen2010:S}
\bibinfo{author}{\bibfnamefont{Y.~L.} \bibnamefont{Chen}},
  \bibinfo{author}{\bibfnamefont{J.-H.} \bibnamefont{Chu}},
  \bibinfo{author}{\bibfnamefont{J.~G.} \bibnamefont{Analytis}},
  \bibinfo{author}{\bibfnamefont{Z.~K.} \bibnamefont{Liu}},
  \bibinfo{author}{\bibfnamefont{K.}~\bibnamefont{Igarashi}},
  \bibinfo{author}{\bibfnamefont{H.-H.} \bibnamefont{Kuo}},
  \bibinfo{author}{\bibfnamefont{X.~L.} \bibnamefont{Qi}},
  \bibinfo{author}{\bibfnamefont{S.~K.} \bibnamefont{Mo}},
  \bibinfo{author}{\bibfnamefont{R.~G.} \bibnamefont{Moore}},
  \bibinfo{author}{\bibfnamefont{D.~H.} \bibnamefont{Lu}},
  \bibnamefont{et~al.}, \bibinfo{journal}{Science}
  \textbf{\bibinfo{volume}{329}}, \bibinfo{pages}{659} (\bibinfo{year}{2010}).

\bibitem[{\citenamefont{Liu et~al.}(2009)\citenamefont{Liu, Liu, Xu, Qi, and
  Zhang}}]{Liu2009:PRL}
\bibinfo{author}{\bibfnamefont{Q.}~\bibnamefont{Liu}},
  \bibinfo{author}{\bibfnamefont{C.-X.} \bibnamefont{Liu}},
  \bibinfo{author}{\bibfnamefont{C.}~\bibnamefont{Xu}},
  \bibinfo{author}{\bibfnamefont{X.-L.} \bibnamefont{Qi}}, \bibnamefont{and}
  \bibinfo{author}{\bibfnamefont{S.-C.} \bibnamefont{Zhang}},
  \bibinfo{journal}{Phys. Rev. Lett.} \textbf{\bibinfo{volume}{102}}
  (\bibinfo{year}{2009}).

\bibitem[{\citenamefont{Hor et~al.}(2010)\citenamefont{Hor, Roushan,
  Beidenkopf, Seo, Qu, Checkelsky, Wray, Hsieh, Xia, Xu et~al.}}]{Hor2010:PRB}
\bibinfo{author}{\bibfnamefont{Y.~S.} \bibnamefont{Hor}},
  \bibinfo{author}{\bibfnamefont{P.}~\bibnamefont{Roushan}},
  \bibinfo{author}{\bibfnamefont{H.}~\bibnamefont{Beidenkopf}},
  \bibinfo{author}{\bibfnamefont{J.}~\bibnamefont{Seo}},
  \bibinfo{author}{\bibfnamefont{D.}~\bibnamefont{Qu}},
  \bibinfo{author}{\bibfnamefont{J.~G.} \bibnamefont{Checkelsky}},
  \bibinfo{author}{\bibfnamefont{L.~A.} \bibnamefont{Wray}},
  \bibinfo{author}{\bibfnamefont{D.}~\bibnamefont{Hsieh}},
  \bibinfo{author}{\bibfnamefont{Y.}~\bibnamefont{Xia}},
  \bibinfo{author}{\bibfnamefont{S.-Y.} \bibnamefont{Xu}},
  \bibnamefont{et~al.}, \bibinfo{journal}{Phys. Rev. B}
  \textbf{\bibinfo{volume}{81}} (\bibinfo{year}{2010}).

\bibitem[{\citenamefont{Ferreira and Loss}(2013)}]{Ferreira2013:PRL}
\bibinfo{author}{\bibfnamefont{G.~J.} \bibnamefont{Ferreira}} \bibnamefont{and}
  \bibinfo{author}{\bibfnamefont{D.}~\bibnamefont{Loss}},
  \bibinfo{journal}{Phys. Rev. Lett.} \textbf{\bibinfo{volume}{111}},
  \bibinfo{pages}{106802} (\bibinfo{year}{2013}).

\bibitem[{\citenamefont{Hammer et~al.}(2013)\citenamefont{Hammer, Ertler, and
  P\"otz}}]{Hammer2013:APL}
\bibinfo{author}{\bibfnamefont{R.}~\bibnamefont{Hammer}},
  \bibinfo{author}{\bibfnamefont{C.}~\bibnamefont{Ertler}}, \bibnamefont{and}
  \bibinfo{author}{\bibfnamefont{W.}~\bibnamefont{P\"otz}},
  \bibinfo{journal}{Appl. Phys. Lett} \textbf{\bibinfo{volume}{102}},
  \bibinfo{pages}{193514} (\bibinfo{year}{2013}).

\bibitem[{\citenamefont{Recher et~al.}(2009)\citenamefont{Recher, Nilsson,
  Burkard, and Trauzettel}}]{Recher2009:PRB}
\bibinfo{author}{\bibfnamefont{P.}~\bibnamefont{Recher}},
  \bibinfo{author}{\bibfnamefont{J.}~\bibnamefont{Nilsson}},
  \bibinfo{author}{\bibfnamefont{G.}~\bibnamefont{Burkard}}, \bibnamefont{and}
  \bibinfo{author}{\bibfnamefont{B.}~\bibnamefont{Trauzettel}},
  \bibinfo{journal}{Phys. Rev. B} \textbf{\bibinfo{volume}{79}},
  \bibinfo{pages}{085407} (\bibinfo{year}{2009}).

\bibitem[{\citenamefont{Pal et~al.}(2011)\citenamefont{Pal, Apel, and
  Schweitzer}}]{Pal2011:PRB}
\bibinfo{author}{\bibfnamefont{G.}~\bibnamefont{Pal}},
  \bibinfo{author}{\bibfnamefont{W.}~\bibnamefont{Apel}}, \bibnamefont{and}
  \bibinfo{author}{\bibfnamefont{L.}~\bibnamefont{Schweitzer}},
  \bibinfo{journal}{Phys. Rev. B} \textbf{\bibinfo{volume}{84}},
  \bibinfo{pages}{075446} (\bibinfo{year}{2011}).

\bibitem[{\citenamefont{Schnez et~al.}(2008)\citenamefont{Schnez, Ensslin,
  Sigrist, and Ihn}}]{Schnez2008:PRB}
\bibinfo{author}{\bibfnamefont{S.}~\bibnamefont{Schnez}},
  \bibinfo{author}{\bibfnamefont{K.}~\bibnamefont{Ensslin}},
  \bibinfo{author}{\bibfnamefont{M.}~\bibnamefont{Sigrist}}, \bibnamefont{and}
  \bibinfo{author}{\bibfnamefont{T.}~\bibnamefont{Ihn}},
  \bibinfo{journal}{Phys. Rev. B} \textbf{\bibinfo{volume}{78}},
  \bibinfo{pages}{195427} (\bibinfo{year}{2008}).

\bibitem[{\citenamefont{Hammer and P\"otz}(2013)}]{Hammer2013:CPC}
\bibinfo{author}{\bibfnamefont{R.}~\bibnamefont{Hammer}} \bibnamefont{and}
  \bibinfo{author}{\bibfnamefont{W.}~\bibnamefont{P\"otz}},
  \bibinfo{journal}{Comput. Phys. Commun. (in press), arXiv:1306.5895}
  (\bibinfo{year}{2013}).

\bibitem[{\citenamefont{Stano and Fabian}(2005)}]{Stano2005:PRB}
\bibinfo{author}{\bibfnamefont{P.}~\bibnamefont{Stano}} \bibnamefont{and}
  \bibinfo{author}{\bibfnamefont{J.}~\bibnamefont{Fabian}},
  \bibinfo{journal}{Phys. Rev. B} \textbf{\bibinfo{volume}{72}},
  \bibinfo{pages}{155410} (\bibinfo{year}{2005}).

\bibitem[{\citenamefont{Kramer et~al.}(2008)\citenamefont{Kramer, Heller, and
  Parrott}}]{Kramer2008:JPCS}
\bibinfo{author}{\bibfnamefont{T.}~\bibnamefont{Kramer}},
  \bibinfo{author}{\bibfnamefont{E.~J.} \bibnamefont{Heller}},
  \bibnamefont{and} \bibinfo{author}{\bibfnamefont{R.~E.}
  \bibnamefont{Parrott}}, \bibinfo{journal}{J. Phys.: Conf. Ser.}
  \textbf{\bibinfo{volume}{99}}, \bibinfo{pages}{012010}
  (\bibinfo{year}{2008}).

\end{thebibliography}

\end{document}